\documentclass[%
reprint,
 superscriptaddress,
 amsmath,amssymb,
 citeautoscript,
 prmaterials,
]{revtex4-2}

\usepackage[version=4]{mhchem}
\usepackage{graphicx}
\usepackage{dcolumn}
\usepackage{bm}
\usepackage{xcolor}

\begin{document}


\title{Site-Specific Structure at Multiple Length Scales in Kagome Quantum Spin Liquid Candidates}

\author{Rebecca W. Smaha}
 \email{rsmaha@stanford.edu}
 \affiliation{Stanford Institute for Materials and Energy Sciences, SLAC National\\Accelerator Laboratory, Menlo Park, California 94025, USA}
 \affiliation{Department of Chemistry, Stanford University, Stanford, California 94305, USA}
 \author{Idris Boukahil}\thanks{These authors contributed equally to this work}
  \affiliation{Department of Physics, Stanford University, Stanford, California 94305, USA}
\affiliation{Theory Institute for Materials and Energy Spectroscopies, SLAC National\\Accelerator Laboratory, Menlo Park, California 94025, USA}
\author{Charles J. Titus}\thanks{These authors contributed equally to this work}
 \affiliation{Department of Physics, Stanford University, Stanford, California 94305, USA}
\author{Jack Mingde Jiang}\thanks{These authors contributed equally to this work}
 \affiliation{Stanford Institute for Materials and Energy Sciences, SLAC National\\Accelerator Laboratory, Menlo Park, California 94025, USA}
 \affiliation{Department of Applied Physics, Stanford University, Stanford, California 94305, USA}
  \author{John P. Sheckelton}
 \affiliation{Stanford Institute for Materials and Energy Sciences, SLAC National\\Accelerator Laboratory, Menlo Park, California 94025, USA}
  \author{Wei He}
 \affiliation{Stanford Institute for Materials and Energy Sciences, SLAC National\\Accelerator Laboratory, Menlo Park, California 94025, USA}
 \affiliation{Department of Materials Science and Engineering, Stanford University, Stanford, California 94305, USA} 
   \author{Jiajia Wen}
\affiliation{Stanford Institute for Materials and Energy Sciences, SLAC National\\Accelerator Laboratory, Menlo Park, California 94025, USA}
\author{John Vinson}
\affiliation{Material Measurement Laboratory, National Institute of Standards and Technology,\\100 Bureau Drive, Gaithersburg, MD 20899}
\author{Suyin Grass Wang}
 \affiliation{NSF's ChemMatCARS, Center for Advanced Radiation Sources, c/o Advanced\\Photon Source/ANL, The University of Chicago, Argonne, Illinois 60439, USA}
\author{Yu-Sheng Chen}
 \affiliation{NSF's ChemMatCARS, Center for Advanced Radiation Sources, c/o Advanced\\Photon Source/ANL, The University of Chicago, Argonne, Illinois 60439, USA}
\author{Simon J. Teat}
 \affiliation{Advanced Light Source, Lawrence Berkeley National Laboratory, Berkeley, California 94720, USA}
 \author{Thomas P. Devereaux}
  \affiliation{Stanford Institute for Materials and Energy Sciences, SLAC National\\Accelerator Laboratory, Menlo Park, California 94025, USA}
   \affiliation{Department of Materials Science and Engineering, Stanford University, Stanford, California 94305, USA} 
 \author{C. Das Pemmaraju}
 \affiliation{Theory Institute for Materials and Energy Spectroscopies, SLAC National\\Accelerator Laboratory, Menlo Park, California 94025, USA}
\author{Young S. Lee}
 \email{youngsl@stanford.edu}
 \affiliation{Stanford Institute for Materials and Energy Sciences, SLAC National\\Accelerator Laboratory, Menlo Park, California 94025, USA}
 \affiliation{Department of Applied Physics, Stanford University, Stanford, California 94305, USA}

\date{\today}

\begin{abstract}
Realizing a quantum spin liquid (QSL) ground state in a real material is a leading issue in condensed matter physics research. In this pursuit, it is crucial to fully characterize the structure and influence of defects, as these can significantly affect the fragile QSL physics. Here, we perform a variety of cutting-edge synchrotron X-ray scattering and spectroscopy techniques, and we advance new methodologies for site-specific diffraction and L-edge Zn absorption spectroscopy. The experimental results along with our first-principles calculations address outstanding questions about the local and long-range structures of the two leading kagome QSL candidates, Zn-substituted barlowite \ce{Cu3Zn$_{x}$Cu$_{1-x}$(OH)6FBr} and herbertsmithite \ce{Cu3Zn(OH)6Cl2}. On all length scales probed, there is no evidence that Zn substitutes onto the kagome layers, thereby preserving the QSL physics of the kagome lattice. Our calculations show that antisite disorder is not energetically favorable and is even less favorable in Zn-barlowite compared to herbertsmithite. 
Site-specific X-ray diffraction measurements of Zn-barlowite reveal that \ce{Cu^2+} and \ce{Zn^2+} selectively occupy distinct interlayer sites, in contrast to herbertsmithite. Using the first measured Zn L-edge inelastic X-ray absorption spectra combined with calculations, we discover a systematic correlation between the loss of inversion symmetry from pseudo-octahedral (herbertsmithite) to trigonal prismatic coordination (Zn-barlowite) with the emergence of a new peak. Overall, our measurements suggest that Zn-barlowite has structural advantages over herbertsmithite that make its magnetic properties closer to an ideal QSL candidate: its kagome layers are highly resistant to nonmagnetic defects while the interlayers can accommodate a higher amount of Zn substitution.
\end{abstract}

\maketitle

\section{\label{sec:Intro}Introduction}

Quantum spin liquid (QSL) materials exhibit an unusual magnetic ground state that is characterized by long-range quantum entanglement of the spins without long-range magnetic order.\cite{Balents2010} While they have been theoretically predicted for many years,\cite{Anderson1973,Anderson1987} experimental breakthroughs have only begun to occur in the past $\approx$15 years.\cite{Shores2005} A promising host for a QSL ground state is the kagome lattice, which consists of corner-sharing triangles. Antiferromagnetic (AF) spins on a kagome lattice exhibit a high level of geometric magnetic frustration; for spin $S=\frac{1}{2}$ nearest-neighbor Heisenberg systems there is no magnetic order, and the ground state is believed to be a QSL.\cite{Sachdev1992,Ran2007,Hermele2008,Jiang2008,Yan2011,Depenbrock2012,Hao2013,He2017} Many of the predicted signatures of QSL physics occur at low energy scales much smaller than the magnetic exchange $J$.

In order to make deep contact between theory and experiment, it has become increasingly clear that a quantitative accounting of the defect structure of the materials must be made. The leading kagome QSL candidate material is the synthetic mineral herbertsmithite, \ce{Cu3Zn(OH)6Cl2}; it has a layered structure crystallizing in rhombohedral space group \textit{R}$\bar{3}$\textit{m} that consists of perfect 2D kagome lattices of $S=\frac{1}{2}$ \ce{Cu^2+} cations separated by non-magnetic \ce{Zn^2+} and \ce{Cl-} ions.\cite{Braithwaite2004,Shores2005,Han2012,Fu2015} The presence and amount of magnetic ``impurities'' of \ce{Cu^2+} between the kagome layers has been the subject of debate, as they obscure measurement of the fundamental physics of the QSL ground state.\cite{Han2016b} Standard crystallographic techniques such as X-ray and neutron diffraction cannot accurately differentiate Cu and Zn as their scattering factors and ionic radii are nearly identical. In order to distinguish site occupancies of these two elements in herbertsmithite, Freedman \textit{et al}. used a novel approach to X-ray anomalous diffraction, in conjunction with extended X-ray absorption fine structure (EXAFS) measurements, to show that up to 15\% \ce{Cu^2+} mixes onto the pseudo-octahedral interlayer \ce{Zn^2+} sites despite best synthetic attempts to substitute a full equivalent of interlayer Zn, leading to a formula of approximately \ce{Cu3Zn$_{0.85}$Cu$_{0.15}$(OH)6Cl2}.\cite{Freedman2010} In addition, it should be highly unlikely for \ce{Zn^2+} to mix onto the kagome sites as these sites are highly distorted (\ce{CuO4Cl2} elongated octahedra). This coordination is consistent with the Jahn-Teller activity of \ce{Cu^2+}, but \ce{Zn^2+} as a Jahn-Teller inactive ion should prefer a more octahedral coordination. Anomalous diffraction and EXAFS confirmed this assignment of negligible mixing.\cite{Freedman2010}

The complications with herbertsmithite have spurred research into related materials in order to develop novel QSL candidates. Another recently discovered mineral with kagome planes of \ce{Cu^2+} is barlowite \ce{Cu4(OH)6FBr}, which crystallizes in hexagonal space group \textit{P}6$_3$/\textit{mmc}.\cite{Elliott2014,Han2014,Jeschke2015,Han2016,Tustain2018,Pasco2018,Smaha2018,Henderson2019}  There are three main structural differences between barlowite and herbertsmithite. 1) The stacking of the kagome planes in barlowite is AA, whereas it is ABCA in herbertsmithite. 2) The interlayer metal coordination is different---barlowite's \ce{Cu^2+}s are disordered over three symmetry-equivalent sites in distorted trigonal prismatic coordination (point group $C_{2v}$). This coordination is relatively rare for \ce{Cu^2+}, although it occurs in several organometallic complexes and related minerals such as the isostructural mineral claringbullite.\cite{Echeverria2009,Burns1995a} However, in herbertsmithite, both the interlayer \ce{Zn^2+} and impurity \ce{Cu^2+} occupy the same centered position (point group $D_{3d}$). 3) At low temperature, crystallographic studies show that herbertsmithite maintains its perfect kagome planes and exhibits no symmetry lowering,\cite{Shores2005,Freedman2010} but barlowite has a clear structural phase transition at $T \approx265$ K to orthorhombic (\textit{Pnma}) or lowered hexagonal symmetry (\textit{P}6$_3$/\textit{m}), depending on the synthesis technique.\cite{Smaha2020} However, standard crystallography yields the average, long-range structure; recent non-crystallographic evidence has pointed to a possible symmetry-lowering local distortion in herbertsmithite,\cite{Zorko2017,Laurita2019,Khuntia2020,Norman2020,Li2020} although this is still under debate.

While in barlowite the \ce{Cu^2+} cations between the kagome layers lead to long-range magnetic order, substituting \ce{Zn^2+} into the compound produces a new QSL candidate, Zn-substituted barlowite \ce{Cu3Zn$_{x}$Cu$_{1-x}$(OH)6FBr}. This is an intriguing material to study as first-principles calculations predict significantly fewer magnetic \ce{Cu^2+} impurities on the interlayer site than in herbertsmithite.\cite{Liu2015a,Guterding2016a} In polycrystalline samples, nearly a full equivalent of Zn can be introduced.\cite{Feng2017,Feng2018,Smaha2018,Smaha2020} We have recently synthesized the first single crystals of Zn-barlowite with no magnetic order with a Zn substitution level of $x \approx 0.5$, and the structure of this compound contains two distinct interlayer sites---the set of three symmetry-equivalent off-center sites observed in barlowite (point group $C_{2v}$) and a centered site (point group $D_{3h}$).\cite{Smaha2020}  This motif is also observed in the rare single crystals found in our previously reported polycrystalline growths\cite{Smaha2020} but has never been observed in herbertsmithite, indicating a fundamental difference in the interlayer behavior between the two compounds. Here, we synthesize deuterated single crystals of Zn-barlowite for the first time, achieving a Zn substitution of $x=0.52$; this will enable future neutron scattering experiments.

To obtain precise information on the presence of defects, we perform synchrotron-based spectroscopic and scattering experiments as well as first-principles simulations on Zn-barlowite and herbertsmithite. We study Zn-barlowite with two distinct levels of Zn substitution, $x\approx0.5$ and $x=0.95$, that both display QSL behavior in contrast to barlowite (the parent compound), which exhibits long-range magnetic order either at $T=15$ K with orthorhombic \textit{Pnma} symmetry or at $T=6$ K with hexagonal \textit{P}6$_3$/\textit{m} symmetry (denoted barlowite \textbf{1} and \textbf{2}, respectively), as described in Ref \cite{Smaha2020}. The results allow us to make important claims regarding the influence of disorder on the QSL behavior.  First, the evidence shows that Zn does not substitute onto the kagome sites. K-edge EXAFS data, a common element-specific technique that measures local structure, are consistent with the lack of anti-site disorder in either herbertsmithite or Zn-barlowite. Site-specific X-ray anomalous diffraction measurements, which probe the long-range crystal structure, definitively determine the Cu/Zn site occupancies of the kagome site, further supporting this conclusion. In addition, they indicate that the two distinct interlayer sites are selectively occupied by Cu or Zn even in highly Zn-substituted barlowite. 

Using powerful new superconducting transition edge sensor detector technology,\cite{tes2019} high-resolution Cu and Zn L-edge X-ray absorption near edge spectroscopy (XANES) measurements of the barlowite family and herbertsmithite are performed, including the first reported Zn L-edge resonant inelastic X-ray scattering (RIXS) data. Comparing the experimental Zn L-edge spectra to first-principles {\sc ocean}~\cite{Vinson_OCEAN1} simulations where Zn has been placed onto either the interlayer site or the kagome site indicates that \ce{Zn^2+} does not occupy the kagome site. Instead, it occupies a centered interlayer site in both Zn-substituted barlowite and herbertsmithite. The simulated and experimental spectra reveal a clear distinction between the local symmetry of the interlayer \ce{Zn^2+} in these two compounds; we show that trigonal prismatic and pseudo-octahedral coordination can be distinguished in Zn L-edge XANES. We find that while the local symmetry of the dilute \ce{Cu^2+} defects are different in Zn-barlowite compared to herbertsmithite, the bulk magnetic properties remain very similar,\cite{Smaha2020} showing that the kagome QSL is robust to this aspect of the material-specific disorder.
 % ...

\section{Methods}\label{methods}

Protonated and deuterated low-temperature orthorhombic barlowite powder (denoted \textbf{1}), low-temperature hexagonal barlowite crystals (denoted \textbf{2}), polycrystalline Zn-substituted barlowite with $x = 0.95$ (denoted \textbf{Zn$^\mathrm{H}_{0.95}$} and \textbf{Zn$^\mathrm{D}_{0.95}$}), and protonated single crystalline Zn-substituted barlowite with $x= 0.56$ (denoted \textbf{Zn$^\mathrm{H}_{0.56}$}) were synthesized as described previously.\cite{Smaha2018,Smaha2020} Polycrystalline and single crystalline herbertsmithite were synthesized as described previously.\cite{Shores2005,Han2011} 

Deuterated single crystalline Zn-substituted barlowite (denoted \textbf{Zn$^\mathrm{D}_{0.52}$}) was prepared by sealing 0.213 g of \ce{CuF2}, 0.796 g of \ce{ZnF2}, and 1.641 g of \ce{LiBr} in a 23 mL PTFE-lined stainless steel autoclave with 15 mL \ce{D2O}. The autoclave was heated over 3 hours to 220 $^{\circ}$C and held for 120 hours, then cooled to 80 $^{\circ}$C over 48 hours. It was held at 80 $^{\circ}$C for 72 hours before being cooled to room temperature over 24 hours. The product was recovered by filtration and washed with DI \ce{H2O}, yielding Zn-barlowite crystals mixed with polycrystalline \ce{LiF}, which was removed by sonication in acetone.

Powder X-ray diffraction (PXRD) data were collected on \textbf{Zn$^\mathrm{D}_{0.52}$} at beamline 11-BM at the Advanced Photon Source (APS), Argonne National Laboratory, at $T = 90$ K and $T=295$ K using an energy of 30 keV. 
Crystalline samples were crushed into a powder and measured in Kapton capillaries. Rietveld refinements were performed using GSAS-II.\cite{Toby2013} Atomic coordinates and isotropic atomic displacement parameters were refined for each atom; site occupancy was also refined for the interlayer site when appropriate. Deuterium was excluded.

Single crystal x-ray diffraction (SCXRD) data sets were collected at $T = 100$ K at NSF's ChemMatCARS beamline 15-ID at the APS using a Bruker APEX II detector and at beamline 12.2.1 at the Advanced Light Source (ALS), Lawrence Berkeley National Laboratory, using a Bruker D85 diffractometer equipped with a Bruker PHOTON II detector. For structure determination, data sets were collected at 30 keV 
for protonated \textbf{Zn$^\mathrm{H}_{0.56}$} (APS) and 17 keV 
for deuterated \textbf{Zn$^\mathrm{D}_{0.52}$} and \textbf{Zn$^\mathrm{D}_{0.95}$} (ALS). The data were integrated and corrected for Lorentz and polarization effects using {\sc saint} and corrected for absorption effects using {\sc sadabs}.\cite{Bruker2016} The structures were solved using intrinsic phasing in {\sc apex3} and refined using the {\sc shelxtl} and {\sc olex2} software.\cite{Bruker2016,Sheldrick2015,Dolomanov2009} Hydrogen atoms were inserted at positions of electron density near the oxygen atom and were refined with a fixed bond length and an isotropic thermal parameter 1.5 times that of the attached oxygen atom. Thermal parameters for all other atoms were refined anisotropically. Further crystallographic information can be found in Ref. \cite{Smaha2020}.

For the anomalous diffraction measurements, fluorescence X-ray absorption spectra (XAS) were collected for each sample near the Cu and Zn K-edges to determine the absorption edge energies. Additional SCXRD data sets were collected on and around the Cu and Zn K-edges (see Table S9). Quantification of the anomalous scattering factors was done using modified software from Ref. \cite{Freedman2010}. The software was updated to support additional symmetry operations as well as fractionally occupied sites, which are both present in our samples. Using the crystal structure solved at high energy, the software refined the wavelength dependent terms of the atomic scattering factor, $f'$ and $f''$, by minimizing $wR_2(F^2)$ against the SCXRD data at each X-ray energy. The software was modified to allow for simultaneous refinement of three atomic sites. In total, seven parameters were refined: an overall structure factor 
and $f'$ and $f''$ for each site (kagome, centered interlayer, and off-center interlayer). The refined values are compared against the calculated values of $f'$ to quantify the degree of Zn-Cu mixing at each site. The calculated values, $f'_c$, are obtained from tabulated data\cite{Sasaki1989} of anomalous scattering factors calculated by Cromer and Liberman's method.\cite{Cromer1970} The relative occupation of species $\alpha$ at site $i$, $p^\alpha_i$, is calculated assuming the measured $f'_i$ is from a linear combination of $f'_c$ of atoms present on that site. For example, to determine Zn occupation on the kagome site, we solve for $p^{Zn}_{kagome}f'^{Zn}_c + (1-p^{Zn}_{kagome})f'^{Cu}_c=f'_{kagome}$.

EXAFS measurements were performed in transmission and fluorescence yield mode at beamline 7-3 at the Stanford Synchrotron Radiation Lightsource (SSRL), SLAC National Accelerator Laboratory. The X-ray beam was monochromatized by a Si(220) double crystal monochromator at $\phi= 0$ orientation and detuned by 50\% to reject harmonics. The beam size was chopped to 1 mm (v) by 10 mm (h) before the sample. Samples were ground to a powder, diluted in \ce{BN}, and packed into sample holders, which were kept at $T = 10$ K during the measurements. Incident and transmitted flux were measured via ionization chambers, and the beam energy was calibrated using the transmission of a Cu or Zn foil downstream from the sample. Fluorescence was detected at 90$^{\circ}$ using a PIPS detector with a $Z-1$ filter (Ni for Cu measurements and Cu for Zn measurements) and Soller slits in order to reduce background fluorescence. The {\sc athena} EXAFS package\cite{Ravel2005} was used to align and calibrate data and fit a spline to the background. Cu measurements were aligned to the first inflection point of the Cu foil, which was assumed to be at 8979 eV. Zn measurements were aligned to the first inflection point of the Zn foil, assumed to be at 9659 eV. The {\sc artemis} package\cite{Ravel2005} was used to fit data using paths calculated by FEFF6.\cite{FEFF} Fluorescence yield spectra were used for all samples except for the Cu EXAFS of polycrystalline samples (barlowite \textbf{1} and \textbf{Zn$^\mathrm{D}_{0.95}$}), which used transmission data.

High resolution Cu and Zn L$_3$-edge XANES measurements were carried out at room temperature at SSRL beamline 10-1.  Samples were pelletized with \ce{BN} or \ce{KBr} and then attached to an aluminum sample holder using carbon tape. Measurements were carried out under high vacuum conditions of $\approx 2.7\times10^{-6}$ Pa ($\approx2\times10^{-8}$ Torr) with a ring current of 500 mA. The synchrotron radiation was monochromatized using the beamline's 1000 line/mm monochromator with entrance and exit slits of 27 $\mu$m. A transition edge sensor (TES) spectrometer\cite{tes2019} was used to collect resonant inelastic X-ray scattering (RIXS) planes with a resolution of 2 eV. The energy measured by the TES was calibrated by periodically measuring a reference sample of graphite, \ce{BN}, \ce{Fe2O3}, \ce{NiO}, \ce{CuO}, and \ce{ZnO}, which provide a stable set of emission lines. From the RIXS planes we extracted the L-$\alpha$ and L-$\beta$ lines, leading to separate L$_2$ and L$_3$ edge partial-fluorescence-yield XAS. 

Theoretical simulations on the Cu and Zn L-edges were performed on barlowite (the high-temperature hexagonal structure in space group \textit{P}6$_3$/\textit{mmc}), idealized \textbf{Zn$_{0.95}$} (\ce{ZnCu3(OH)6FBr}), and idealized herbertsmithite (\ce{ZnCu3(OH)6Cl2}) using the crystallographically-determined structures\cite{Freedman2010,Smaha2018,Smaha2020} as a starting point. To simplify the calculations for \textbf{Zn$_{0.95}$} and herbertsmithite, full occupancy of \ce{Zn^2+} on the respective centered interlayer site of each compound and full occupancy of \ce{Cu^2+} on the kagome sites were used. For barlowite, a single off-center ($C_{2v}$) configuration was chosen for the interlayer \ce{Cu^2+} cation, ignoring the site occupancy disorder. Upon structural relaxation, the interlayer \ce{Cu^2+} kept its off-center position. Models using Zn substituted on a kagome site (stoichiometrically equivalent to the interlayer Zn models) were constructed such that there is one \ce{Zn^2+} per kagome layer. Structural relaxations were performed using density functional theory (DFT) with projector-augmented wave pseudopotentials as implemented in the Vienna \textit{ab initio} simulation package (VASP) code.\cite{kresse1993ab,kresse1994ab,kresse1996efficiency,kresse1996efficient} The PBEsol functional \cite{pbesol} was employed with an on-site Coulomb repulsion $U$ set to 5 eV to account for electronic correlations in Mott-Hubbard systems. All degrees of freedom were allowed to relax until the change in energy per ion was less than $1 \times 10^{-5}$ eV using a $4 \times 4 \times 4$ $\Gamma$-centered k-point sampling mesh and a plane wave cut-off energy of 520 eV.

The {\sc ocean} package,\cite{Vinson_OCEAN1,Gilmore_OCEAN2} which solves the Bethe-Salpeter equation (BSE) for core-level excitations, was employed to calculate the Cu and Zn L-edge spectra. The BSE is solved in conjunction with the ground state electronic structure obtained using the local density approximation (LDA) in the plane wave basis DFT code Quantum Espresso.\cite{QE_Giannozzi} These calculations used norm conserving pseudopotentials treating O:$2s,2p$, F:$2s,2p$, Cl:$3s,3p$, Cu:[Ar]$4s^1 3d^{10}$, Zn:[Ar]$4s,3d$, and Br:$4s,4p$ states in the valence with a plane wave cutoff of 100 Ry. Optimized geometries obtained using the VASP code as described above were used in these simulations. Convergence was achieved with a $6 \times 6 \times 6$ k-point sampling mesh for the final state wavefunctions and a $2 \times 2 \times 2$ grid for the screened core-hole interaction. The theoretical spectra were energy aligned to the first peak in the corresponding experimental spectrum.

\section{Results}
\subsection{Synthesis and Crystal Structure}\label{syn}

\begin{figure}
\includegraphics[width=8cm]{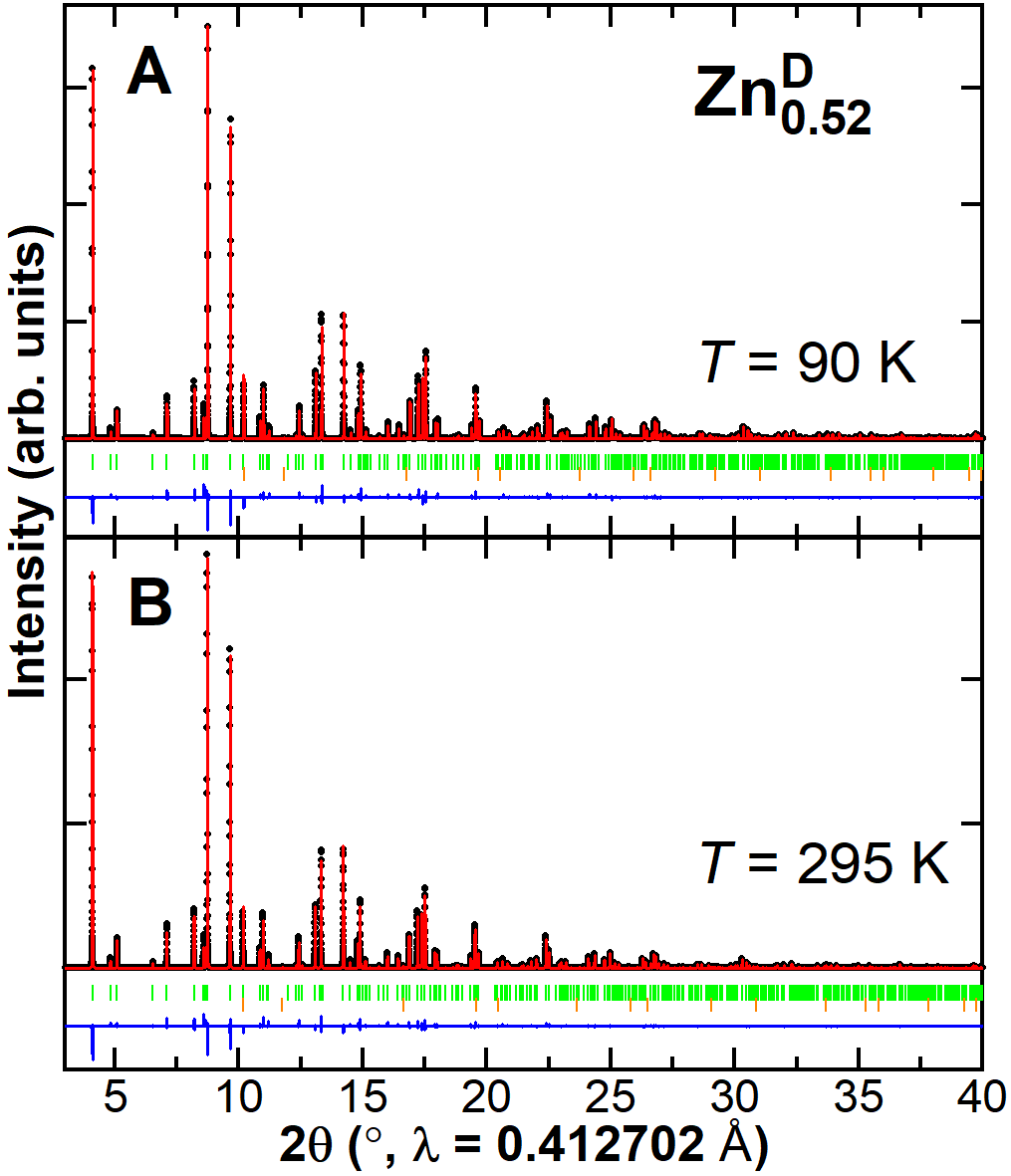}
\caption{Rietveld refinements of synchrotron PXRD data of deuterated \textbf{Zn$^\mathrm{D}_{0.52}$} at A) $T = 90$ K and B) $T=295$ K  in space group \textit{P}6$_3$/\textit{mmc}. Observed (black), calculated (red), and difference (blue) plots are shown, and Bragg reflections are indicated by green tick marks. The Bragg reflections of a \ce{LiF} impurity phase are marked with orange tick marks.}
\label{fgr:Riet}
\end{figure}

In this work, we report the first synthesis of deuterated single crystals of Zn-substituted barlowite, achieving a Zn substitution of $x=0.52$ (\ce{Cu$_{3.48}$Zn$_{0.52}$(OD)6FBr}) as measured by inductively coupled plasma atomic emission spectroscopy (ICP-AES); the availability of these crystals will enable future neutron scattering experiments of the ground state physics of this compound.  This sample is denoted \textbf{Zn$^\mathrm{D}_{0.52}$}, where the superscript indicates if it is protonated or deuterated and the subscript indicates the level of Zn substitution.  Rietveld refinements of synchrotron powder X-ray diffraction (PXRD) data of \textbf{Zn$^\mathrm{D}_{0.52}$} collected at $T = 90$ and $295$ K are shown in Figure \ref{fgr:Riet}, and synchrotron single crystal x-ray diffraction (SCXRD) was performed at $T = 100$ K. Crystallographic data from these measurements are tabulated in Tables S1--S6 in the Supplemental Material.\cite{supp}  The average formula from all diffraction measurements is \ce{Cu$_{3.54}$Zn$_{0.46}$(OH)6FBr}, consistent with the ICP-AES results. 

We also investigate samples reported previously: two compositions of Zn-substituted barlowite with different levels of Zn substitution synthesized via distinct chemical reactions.\cite{Smaha2018,Smaha2020} Protonated single crystals analogous to the deuterated crystals reported here achieved a maximum Zn substitution of $x =0.56$. This yielded the formula \ce{Cu$_{3.44}$Zn$_{0.56}$(OH)6FBr} (denoted \textbf{Zn$^\mathrm{H}_{0.56}$}); detailed crystallographic studies are reported in Ref. \cite{Smaha2020}. Polycrystalline samples with $x = 0.95$ are denoted \textbf{Zn$_{0.95}$}; this is comparable to---but higher than---the level of substitution achieved in herbertsmithite.\cite{Freedman2010} Its structure was determined by synchrotron PXRD and neutron powder diffraction (NPD).\cite{Smaha2020}  A crystal, one of fewer than 15 found in the synthesis of nearly 20 g of the polycrystalline sample, was measured via synchrotron SCXRD in Ref. \cite{Smaha2020}; this is the \textbf{Zn$^\mathrm{D}_{0.95}$} sample upon which X-ray anomalous diffraction measurements are performed here, as described in Section \ref{anom}.  It has non-negligible electron density on the triplicated off-center ($C_{2v}$) interlayer sites---approximately 5\% on each, for a total of 15\%---leading to an empirical formula (assuming Cu on the kagome and off-center interlayer sites and Zn on the centered interlayer site) of \ce{Cu$_{3.15}$Zn$_{0.85}$(OD)6FBr}. This is slightly different from the bulk formula determined by ICP-AES of \ce{Cu$_{3.05}$Zn$_{0.95}$(OD)6FBr}.\cite{Smaha2020}  Both facts suggest that this crystal may not be representative of the bulk. However, it is valuable to test whether the structural trends observed in the samples with $x\approx0.5$ are robust to higher Zn substitution values, and it allows a structural comparison with herbertsmithite, which has approximately the same amount of Zn substitution.

Both the $x = 0.95$ and $x \approx 0.5$ compositions of Zn-barlowite crystallize in space group \textit{P}6$_3$/\textit{mmc} (No. 194) down to the lowest measured temperatures; this lack of symmetry lowering implies that their kagome lattices remain perfect and undistorted. SCXRD refinements (from this work for \textbf{Zn$^\mathrm{D}_{0.52}$} and Ref. \cite{Smaha2020} for \textbf{Zn$^\mathrm{H}_{0.56}$} and \textbf{Zn$_{0.95}$}) indicate that both have two distinct sites for the interlayer metal ions (see Figure \ref{fgr:schematic}), but the similarity of the X-ray scattering factors of Cu and Zn makes it impossible to distinguish these elements accurately. Jahn-Teller theory provides a reasonable guess that \ce{Cu^2+} should occupy the kagome site (which has heavily elongated \ce{CuO4Br2} octahedra) and the set of three off-center, distorted trigonal prismatic ($C_{2v}$) interlayer sites observed in all-Cu barlowite.\cite{Echeverria2009}  The coordination of each off-center interlayer site consists of four short and two long \textit{M}--O bonds ($\approx$2.0 and 2.4 \AA, respectively). In contrast, the Jahn-Teller inactivity of \ce{Zn^2+} should predispose it to occupy the centered interlayer site in $D_{3h}$ point group symmetry, which has 6 equivalent \textit{M}--O bond lengths ($\approx$2.1 \AA).

\subsection{EXAFS Measurements}
Extended X-ray absorption fine structure (EXAFS) measurements at the Cu and Zn K-edges provide element-specific insight into the local geometry, in contrast to crystallography which gives a global picture of long-range crystalline order. Crystallographic data from Ref. \cite{Smaha2020} were used to generate the base paths and coordination numbers for the EXAFS fits. It was assumed that the kagome sites are fully occupied by Cu and that in Zn-substituted barlowite the off-center interlayer site is occupied by Cu while the centered interlayer site is occupied by Zn; further experimental validation of these assumptions will be discussed in Section \ref{anom}. Fit details are discussed at greater length in the Supplemental Material and shown in Tables S7--S8. The real parts of $\chi(R)$ are shown in Figure S1. 

\begin{figure}
\includegraphics[width=8cm]{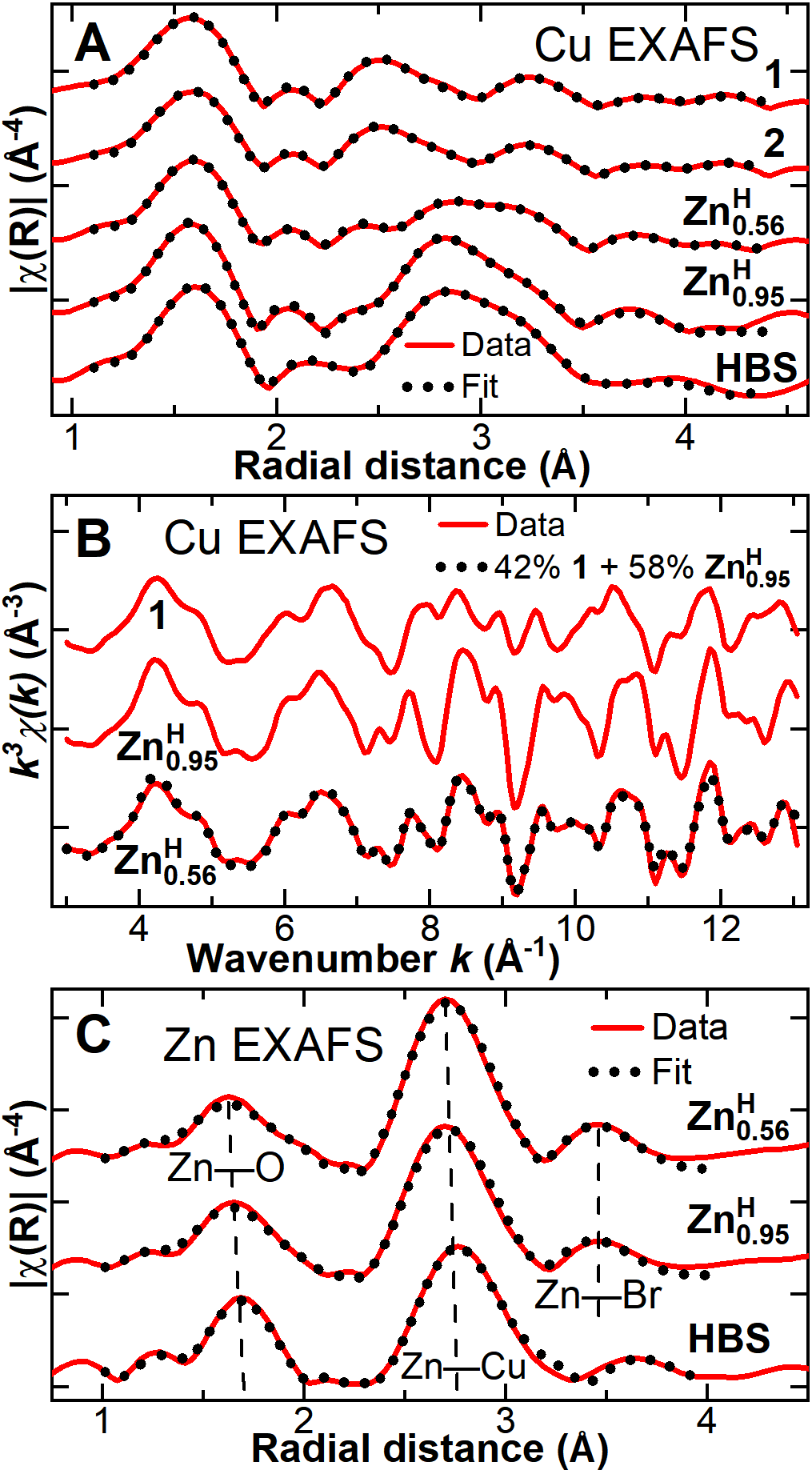}
\caption{A) Real-space Cu K-edge EXAFS data and fits of barlowite \textbf{1}, barlowite \textbf{2}, \textbf{Zn$^\mathrm{H}_{0.56}$}, \textbf{Zn$^\mathrm{H}_{0.95}$}, and herbertsmithite (HBS) showing the magnitude of the Fourier-transformed EXAFS ($\chi(R)$) measured at $T = 10$ K. B) Linear component fit of normalized \textbf{Zn$^\mathrm{H}_{0.56}$} Cu EXAFS data of barlowite \textbf{1} and \textbf{Zn$^\mathrm{H}_{0.95}$}. C) Real-space Zn K-edge EXAFS data and fits to  \textbf{Zn$^\mathrm{H}_{0.56}$}, \textbf{Zn$^\mathrm{H}_{0.95}$}, and herbertsmithite (HBS) measured at $T = 10$ K. The three paths for Zn-substituted barlowite are shown as dashed lines. Vertical offsets have been applied to separate the spectra.}
\label{fgr:exafs}
\end{figure}

Cu EXAFS data and fits for barlowite \textbf{1} and \textbf{2}, Zn-barlowite \textbf{Zn$^\mathrm{H}_{0.56}$} and \textbf{Zn$^\mathrm{H}_{0.95}$}, and herbertsmithite are shown in Figure \ref{fgr:exafs}A. Our EXAFS models reproduce the data well, capturing all significant features and confirming that the local structure is consistent with the crystallographic data. However, in contrast to the crystallography that reveals different symmetries for \textbf{1} and \textbf{2} at low temperatures (orthorhombic \textit{Pnma} and hexagonal \textit{P}6$_3$/\textit{m}, respectively), their EXAFS spectra are identical, suggesting that the global symmetry differences of the long-range structures between these samples are not tied to changes in the local environment of the Cu. The differences between the \textbf{Zn$^\mathrm{H}_{0.95}$} and herbertsmithite spectra around $\approx$2.25 \AA\ are due to the different bond length of Cu--Br compared to Cu--Cl. In Figure \ref{fgr:exafs}B, a linear component analysis of normalized Cu EXAFS data for \textbf{Zn$^\mathrm{H}_{0.56}$} is shown, confirming \textbf{Zn$^\mathrm{H}_{0.56}$} as a midpoint between barlowite \textbf{1} and \textbf{Zn$^\mathrm{H}_{0.95}$}.  The \textbf{Zn$^\mathrm{H}_{0.56}$} data are fit well by the expected blend of barlowite \textbf{1} and \textbf{Zn$^\mathrm{H}_{0.95}$}, which indicates 42\% Cu on the interlayer, consistent with the stoichiometry found via ICP-AES and the relative occupancies of the two types of interlayer sites observed crystallographically. 

Figure \ref{fgr:exafs}C shows fits to Zn K-edge EXAFS data for \textbf{Zn$^\mathrm{H}_{0.56}$}, \textbf{Zn$^\mathrm{H}_{0.95}$}, and herbertsmithite; they are consistent with locating all Zn on the interlayer. Both Zn-substituted barlowite samples can be fit well with the Zn--O, Zn--Cu, and Zn--Br bond lengths expected for interlayer \ce{Zn^2+}. Were \ce{Zn^2+} to occupy the kagome site, the paths should differ primarily via the inclusion of a 3.3 \AA\ Zn--Cu bond; attempting this does not improve the fit of either \textbf{Zn$^\mathrm{H}_{0.56}$} or \textbf{Zn$^\mathrm{H}_{0.95}$}, and when the occupancy of this site is allowed to freely refine, it is indistinguishable from zero. In addition, our herbertsmithite Zn K-edge data are in broad agreement with a previous report, which concluded that no statistically significant amount of \ce{Zn^2+} occupies the kagome layer.\cite{Freedman2010} The spectra of \textbf{Zn$^\mathrm{H}_{0.56}$} and \textbf{Zn$^\mathrm{H}_{0.95}$} are nearly identical, indicating that \ce{Zn^2+} occupies the same site in both compositions.

While EXAFS paints a consistent picture of the local environment in barlowite and herbertsmithite, this approach has inherent limitations. Although EXAFS is an element-specific technique, it is not site-specific, requiring a model which averages over all sites. Within this framework, it is easy to conclude that the Zn spectra can be fit well by a model that only includes an interlayer site, but difficult to conclusively determine that \ce{Zn^2+} is not present on the kagome site with some small occupancy, as this signal would represent a small average contribution to the EXAFS spectrum. Furthermore, the crystalline nature of these samples results in a large number of single-scattering paths---so many that it was not possible to include them all in a model and still respect the degree-of-freedom constraint imposed by the measured $k$-range.\cite{Newville2014} This statistical constraint prevents the application of even more complex models. Thus, although we have demonstrated that the EXAFS of Zn-barlowite is consistent with no occupation of the kagome site by \ce{Zn^2+} and improved upon the measurements of herbertsmithite originally performed by Ref. \cite{Freedman2010}, an element-specific technique is not sufficient. Resolving conclusively which crystallographic sites within Zn-substituted barlowite are occupied by \ce{Cu^2+} and \ce{Zn^2+} requires a technique that is both element-specific \emph{and} site-specific, as discussed below.

\subsection{X-Ray Anomalous Diffraction}\label{anom}
We performed multi-wavelength X-ray anomalous diffraction measurements to provide both element specificity and site specificity to characterize the Cu/Zn populations on kagome and interlayer sites. Similar measurements were previously employed to successfully distinguish Cu/Zn occupancy in herbertsmithite.\cite{Freedman2010} For compounds containing elements of similar $Z$, their atomic form factors can be indistinguishable, and therefore any site mixing between such elements cannot be effectively resolved with conventional crystallography. However, the anomalous dispersion factors $f'$ and $f''$ for each atomic species vary sensitively near absorption edges with changing incident X-ray energy. Combined with crystallography, one can determine the structure and elemental populations at specific crystallographic sites.  

\begin{figure}[ht]
\includegraphics[width=8cm]{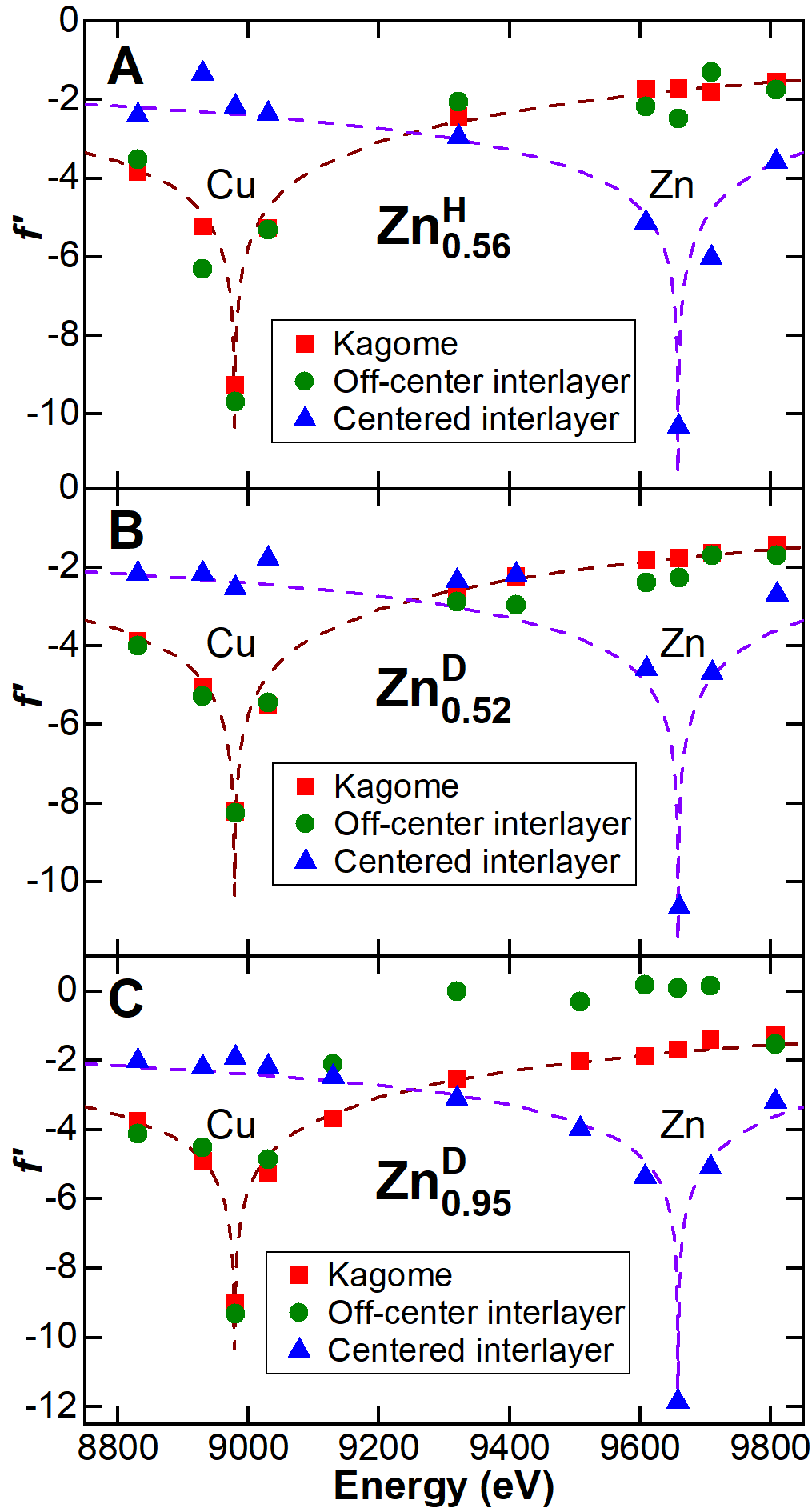}
\caption{Anomalous dispersion factor $f'$ as a function of incident energy for the kagome, off-center interlayer ($C_{2v}$), and centered interlayer ($D_{3h}$)  sites in A) protonated \textbf{Zn$^\mathrm{H}_{0.56}$}, B) deuterated \textbf{Zn$^\mathrm{D}_{0.52}$}, and C) deuterated \textbf{Zn$^\mathrm{D}_{0.95}$}. The theoretical values for Cu and Zn are shown as maroon and purple dashed lines, respectively.}
\label{fgr:anom}
\end{figure}

We have significantly broadened the analysis capabilities of software previously developed specifically for herbertsmithite,\cite{Freedman2010} which supported a subset of the possible symmetry operations and allowed the co-refinement of only two crystallographic sites. While the measurements require a synchrotron to tune the incident radiation and may be time consuming, anomalous diffraction is a powerful technique with potential for broad applications, particularly for distinguishing site mixing associated with doping/substitution that is commonly found in solid state compounds and condensed matter systems.

Synchrotron SCXRD data sets were collected on protonated \textbf{Zn$^\mathrm{H}_{0.56}$} and deuterated \textbf{Zn$^\mathrm{D}_{0.52}$} and \textbf{Zn$^\mathrm{D}_{0.95}$} at $T = 100$ K with energies selected around the absorption K-edges of Cu and Zn and at high energy far from these absorption edges (tabulated in Table S9). We refined $f'$ and $f''$ for each site that may contain Cu or Zn (kagome, centered interlayer, and off-center interlayer) against crystal structures collected at high energy, as shown in Figure \ref{fgr:anom}. For all three samples, $f'$ shows no appreciable decline at the Zn edge for the kagome site, nor at the Cu edge for the centered interlayer site within experimental uncertainty. This demonstrates that there is no measurable \ce{Zn^2+} mixing on the kagome site nor \ce{Cu^2+} mixing on the centered interlayer ($D_{3h}$) site, consistent with Jahn-Teller theory.

\begin{figure}
\includegraphics[width=8cm]{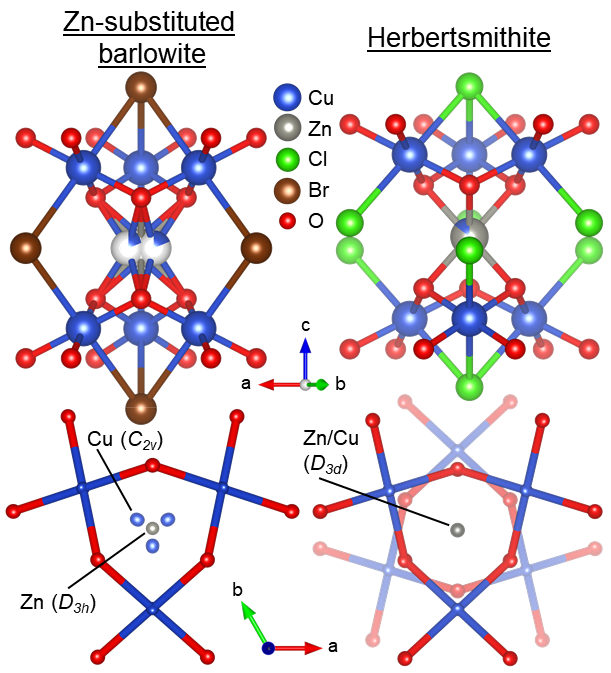}
\caption{Schematic comparing the structures and interlayer coordination of Zn-substituted barlowite (left) and herbertsmithite (right), visualized in VESTA.\cite{Momma2011} Top: side-on views showing the kagome \ce{Cu^2+}s' elongated \ce{CuO4Br2} or \ce{CuO4Cl2} octahedra and the interlayer sites. F and H (or D) atoms are not shown.  Bottom: views of the interlayer metal coordination down the \textit{c}-axis. The atoms are displayed as thermal ellipsoids at 90\% probability.  F, Br, Cl, and H (or D) atoms are not shown. We display the SCXRD structure of \textbf{Zn$^\mathrm{D}_{0.52}$} reported in this work and the herbertsmithite structure reported by Ref. \cite{Freedman2010}.}
\label{fgr:schematic}
\end{figure}

The centered interlayer site is occupied by Zn in all samples. In \textbf{Zn$^\mathrm{H}_{0.56}$} and \textbf{Zn$^\mathrm{D}_{0.52}$} (Figure \ref{fgr:anom}A and B), the slight dips of $f'$ at the Zn edge for the off-center interlayer sites (in distorted trigonal prismatic coordination) puts the upper bound of Zn occupancy at $\approx$8\% and $\approx$7\%, respectively.  In the single crystal found in the deuterated \textbf{Zn$^\mathrm{D}_{0.95}$} sample, the small atomic occupancy on each of the three off-center interlayer sites (approximately 5\%\cite{Smaha2020}) contributes to large errors in $f'$ away from elemental edges, as shown in Figure \ref{fgr:anom}C. The lack of any noticeable dip in $f'$ on the Zn edge for the off-center interlayer sites indicates negligible presence of Zn on these sites for all three samples. 

Prior Rietveld co-refinements of NPD and synchrotron PXRD data of \textbf{Zn$^\mathrm{D}_{0.95}$}  produce good fits using the site assignments of Zn and Cu extracted from these anomalous diffraction results.\cite{Smaha2020}  Although ICP-AES indicates that there should be approximately $5\%$ interlayer Cu in \textbf{Zn$^\mathrm{D}_{0.95}$},  it will be disordered over the three symmetry-equivalent off-center interlayer sites, leading to a mere $\approx$1.5\% site occupancy on each site---this is nearing the detection limit of the powder diffraction data. Thus, we used a Rietveld co-refinement model for \textbf{Zn$^\mathrm{D}_{0.95}$} containing only a centered interlayer site fully occupied with Zn.  The Rietveld refinements of PXRD data of protonated \textbf{Zn$^\mathrm{H}_{0.56}$} (in Ref. \cite{Smaha2020}) and deuterated \textbf{Zn$^\mathrm{D}_{0.52}$} (in Section \ref{syn}) also utilize the model extracted from these anomalous diffraction results, in which the kagome site and off-center interlayer sites are fully occupied by \ce{Cu^2+} and the centered interlayer site is fully occupied by \ce{Zn^2+}.  Cu/Zn mixing is not included in the model, consistent with the low degree of likelihood that any site is simultaneously occupied by both \ce{Cu^2+} and \ce{Zn^2+}, as indicated in Figure \ref{fgr:anom}.

The clear existence of two types of interlayer sites in all Zn-substituted barlowite samples, and the selective occupation (consistent with Jahn-Teller predictions) of the centered and off-center sites by \ce{Zn^2+} and \ce{Cu^2+}, respectively, invite comparison with herbertsmithite.  Since SCXRD was able to detect the presence of the off-center interlayer sites at $\approx$5\% site occupancy in the single crystal found in the bulk \textbf{Zn$^\mathrm{D}_{0.95}$} sample, it should be able to detect similarly occupied sites in herbertsmithite. However, in herbertsmithite, only one interlayer site has been observed in all crystallographic studies (including synchrotron SCXRD, synchrotron PXRD, and NPD), and this site is occupied by a mixture of approximately 85\% \ce{Zn^2+} and 15\% \ce{Cu^2+}.\cite{Shores2005,Freedman2010,Chu2011,Han2011,Welch2014} Figure \ref{fgr:schematic} depicts the different interlayer coordination environments of Zn-barlowite and herbertsmithite in two views: side-on (in the \textit{ac} plane) and top-down (in the \textit{ab} plane). To further understand the nature of the differences in local coordination on the interlayer site(s) between Zn-substituted barlowite and herbertsmithite, we turned to theoretical simulations of the electronic structure.

\subsection{Simulated and Experimental XANES Measurements}

\begin{figure}
\includegraphics[width=8cm]{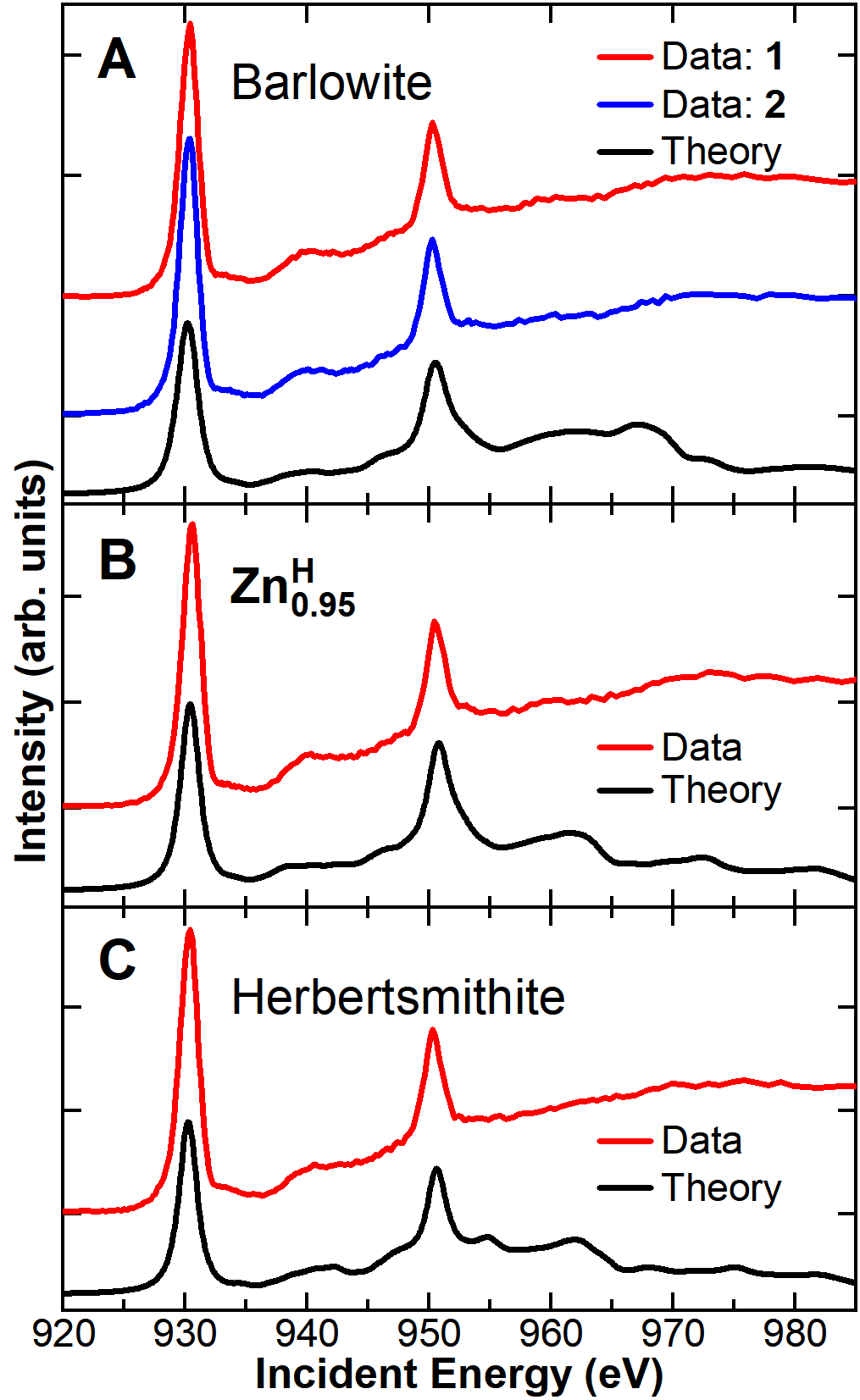}
\caption{Comparison of room temperature experimental Cu L-edge XANES spectra with simulations for A) barlowite \textbf{1} and \textbf{2}, B) \textbf{Zn$^\mathrm{H}_{0.95}$}, and C) herbertsmithite.  Vertical offsets have been applied to separate the spectra. To match the experimental spectra, the calculated spectra were broadened with a 0.6 eV Gaussian and a 0.8 eV Lorentzian broadening. }
\label{fgr:CuXAS}
\end{figure}

\begin{figure*}
\includegraphics[width=16cm]{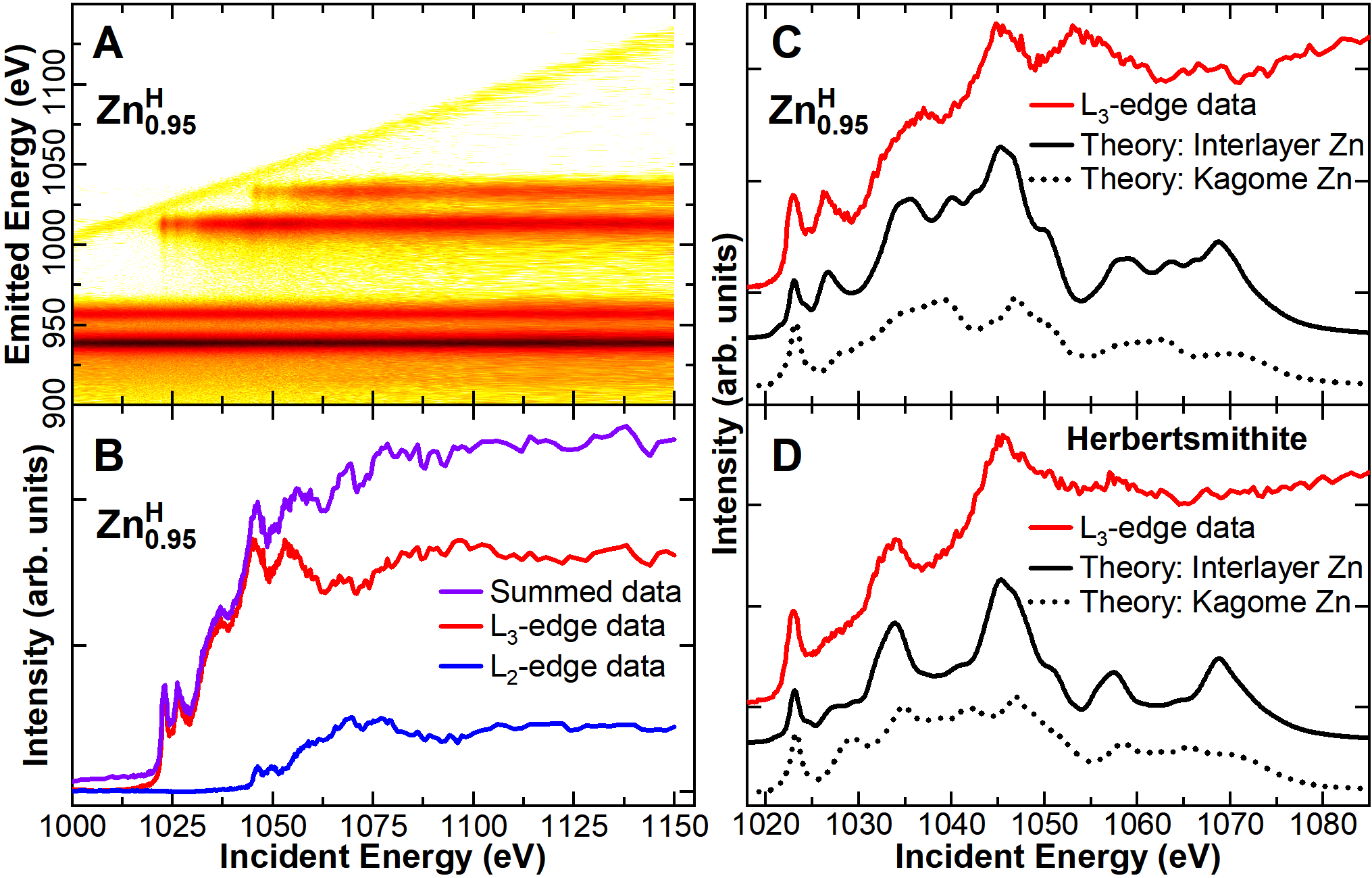}
\caption{ A) Zn L-edge RIXS plane of \textbf{Zn$^\mathrm{H}_{0.95}$} showing the L-$\alpha$ fluorescence with emitted energy of 1012 eV and the L-$\beta$ fluorescence with emitted energy of 1035 eV, corresponding to the L$_3$ and L$_2$ edges, respectively. Fluorescence from the Cu L-$\alpha$ and L-$\beta$ lines are visible between 920 and 960 eV, and the elastic scattering is visible as a diagonal line. A log scale has been used on the color map to emphasize the faint elastic line. 
B) Zn L-edge XANES data of \textbf{Zn$^\mathrm{H}_{0.95}$} measured at $T = 300$ K. The TES detector allows the L$_2$-edge and L$_3$-edge data to be separated; they are shown along with the summed data. Zn L$_3$-edge XANES data of C) \textbf{Zn$^\mathrm{H}_{0.95}$} and D)  herbertsmithite measured at $T = 300$ K, along with simulations based on the experimentally determined crystal structure (``interlayer Zn'') and simulations where Zn occupies one of the kagome sites while Cu occupies the respective interlayer site in each structure (``kagome Zn''; see discussion in text). In C) and D), vertical offsets have been applied to separate the spectra. The calculated Zn spectra were broadened by convolution with a 0.8 eV Gaussian broadening, and the non-excitonic portion had an additional 0.8 eV Lorentzian broadening. }
\label{fgr:ZnXAS}
\end{figure*}

Spectra based on idealized versions of the experimental crystal structures of room-temperature (\textit{P}6$_3$/\textit{mmc}) barlowite, \textbf{Zn$_{0.95}$}, and herbertsmithite (as described in Section \ref{methods}) were calculated using the {sc ocean} package and are shown in Figures \ref{fgr:CuXAS} and \ref{fgr:ZnXAS} for the Cu and Zn L-edges, respectively, where they are referred to as the ``interlayer Zn'' model. To experimentally verify the simulations, we measured room temperature high-resolution X-ray absorption near edge spectra (XANES) at the Cu and Zn L-edges, which probe element-specific symmetry and electronic structure. The Cu and Zn L-edge spectra both unambiguously determine a 2+ valency, as expected.  The simulations and experimental Cu L-edge spectra for barlowite, \textbf{Zn$^\mathrm{H}_{0.95}$}, and herbertsmithite, in Figure \ref{fgr:CuXAS}, are in good agreement for all samples. The experimental Cu L-edge spectrum of \textbf{Zn$^\mathrm{H}_{0.56}$} is shown in Figure S2A.

We present here the first Zn L-edge absorption data collected with a transition edge sensor (TES) detector; it collects resonant inelastic X-ray scattering (RIXS) planes, allowing for the L$_3$-edge and L$_2$-edge spectra to be easily separated and measured with extremely low noise.  To the extent of our knowledge, this is also the first published Zn L-edge RIXS data. Without the high energy resolution afforded by the TES, residual fluctuations from the Cu edge would overwhelm the Zn signal, making it essential for these samples in particular (as illustrated in Figure S3). The measured Zn L-edge RIXS plane of \textbf{Zn$^\mathrm{H}_{0.95}$} is depicted in Figure \ref{fgr:ZnXAS}A, showing the Cu fluorescent tail and elastic scattering. An example of the two Zn L-edge components (derived from the RIXS plane) and their sum, which is what lower-resolution detectors measure, is plotted in Figure \ref{fgr:ZnXAS}B for \textbf{Zn$^\mathrm{H}_{0.95}$}. The analogous separated spectra for \textbf{Zn$^\mathrm{H}_{0.56}$} and herbertsmithite are plotted in Figure S2. The calculated partial density of states (DOS) of idealized barlowite (room temperature), \textbf{Zn$_{0.95}$}, and herbertsmithite, shown in Figures S4--S5, are consistent with previous results and similar across these three compounds.\cite{Jeschke2013,Jeschke2015,Liu2018,Jiang2018}

The L-edge XAS of \ce{Zn^2+} is inherently more sensitive to local structure than \ce{Cu^2+}. In \ce{Cu^2+}, the absorption spectrum is dominated by strong white-line transitions into the unoccupied $3d$ state, which is relatively unaffected by coordination and bond strength. On the other hand, in \ce{Zn^2+} the $3d$ orbitals are occupied and the transitions are to the more diffuse $4s$ orbitals.  Moreover, \ce{Zn^2+} occupies a unique position within the structures of both Zn-substituted barlowite and herbertsmithite, allowing us to precisely compare the coordination environments of the interlayer sites between these two compounds.  The measured Zn L-edge XANES spectra display good agreement with the simulated ``interlayer Zn'' spectra, as shown in Figure \ref{fgr:ZnXAS}C and D. Just as in the Zn K-edge EXAFS data, the Zn L-edge XANES of \textbf{Zn$^\mathrm{H}_{0.56}$} (Figure S2B) and \textbf{Zn$^\mathrm{H}_{0.95}$} match well, indicating that their local coordinations are identical; this confirms that the \ce{Zn^2+} occupies the same site in both compositions. Calculations were also performed for modified unit cells of idealized \textbf{Zn$_{0.95}$} and herbertsmithite with \ce{Zn^2+} placed onto a kagome site and \ce{Cu^2+} placed onto the interlayer site; this model is referred to as ``kagome Zn'' in Figure \ref{fgr:ZnXAS}. These simulated spectra deviate significantly from the experimental spectra and the ``interlayer Zn'' models. The mismatch between the experimental data and ``kagome Zn'' model for \textbf{Zn$^\mathrm{H}_{0.95}$} (Figure \ref{fgr:ZnXAS}C) is particularly visible in the 1025--1030 eV region, where the ``kagome Zn'' spectrum is missing a peak that is clearly present in the experimental data.  The mismatch is reversed in the herbertsmithite spectra (Figure \ref{fgr:ZnXAS}D), where the ``kagome Zn'' simulated spectrum has more of a peak in this region than the experimental spectrum.  These simulations illustrate that no significant amount of antisite (Zn-on-kagome) disorder occurs in either Zn-substituted barlowite and herbertsmithite.

Moreover, the free energies of the relaxed ``kagome Zn'' structures are significantly less stable than those of the relaxed ``interlayer Zn'' structures. For herbertsmithite, the ``interlayer Zn'' model is more stable than the ``kagome Zn'' model by 0.25 
eV per Zn, while for Zn-barlowite the ``interlayer Zn'' model is more stable by 0.44 
 eV per Zn. This provides a thermodynamic rationale which further confirms that it is highly unlikely that \ce{Zn^2+} cations substitute onto the kagome lattice. Additionally, the difference between the compounds suggests that Zn-barlowite is more stable than herbertsmithite and less susceptible to Zn-on-kagome antisite disorder.

While the Zn L-edge XANES spectra of all samples have a strong peak at $\approx$1023 eV, both Zn-barlowites (\textbf{Zn$^\mathrm{H}_{0.95}$} and \textbf{Zn$^\mathrm{H}_{0.56}$} in Figures \ref{fgr:ZnXAS}A--C and S2B) have a second strong peak at $\approx$1026.5 eV that is significantly suppressed in the herbertsmithite spectra (Figure \ref{fgr:ZnXAS}D). These data illustrate a fundamental difference in the interlayer \ce{Zn^2+}'s local coordination between Zn-substituted barlowite and herbertsmithite, as illustrated in Figure \ref{fgr:schematic}.  While both have six equivalent Zn--O bonds where the O's form two equilateral triangles above and below the Zn, in herbertsmithite the two triangles are staggered, while in Zn-barlowite the triangles are eclipsed.  Thus, the local environment of \ce{Zn^2+} in herbertsmithite is pseudo-octahedral and has inversion symmetry (point group $D_{3d}$).\cite{Echeverria2009} Conversely, the \ce{Zn^2+} in Zn-barlowite has trigonal prismatic local coordination (point group $D_{3h}$), with a loss of inversion symmetry. While this coordination for \ce{Zn^2+} is rare, it is not unprecedented; in a statistical survey of oxides, trigonal prismatic local coordination was found in 1.23\% of Zn sites.\cite{Waroquiers2017} Calculated electron density isosurfaces confirm that both Zn-barlowite and herbertsmithite have peaks at $\approx$1023 eV but differ substantially at the energy of the peak observed only in Zn-barlowite ($\approx$1026.5 eV; see Figure S6). To check if this effect is systematic, we performed analogous Zn L-edge measurements and simulations on materials with similar \ce{Zn^2+} coordination environments (pseudo-octahedral rutile \ce{ZnF2} and trigonal prismatic \ce{Sr3ZnRhO6});\cite{Layland2000} as shown in Figures S7--S9, they exhibit similar spectra to herbertsmithite and Zn-barlowite, respectively. Thus, the simple one-peak spectrum appears to be characteristic of 6-coordinate \ce{Zn^2+} with a locally cubic or pseudo-cubic/pseudo-octahedral environment, and the second prominent peak observed in Zn-barlowite results from the trigonal prismatic coordination and the subsequent loss of inversion symmetry. 

\section{\label{sec:Discussion}Discussion and Conclusion}

In this work, we investigate several outstanding questions about the structure and possible site-mixing disorder in Zn-substituted barlowite and compare it to the best known kagome QSL candidate material, herbertsmithite.  We first probe the likelihood of any \ce{Zn^2+} substituting onto the kagome site, as it is desirable for the lattice to be as ideal and undistorted as possible. This question cannot be answered using standard crystallographic techniques (X-ray diffraction and powder neutron diffraction) due to the nearly identical scattering factors of Cu and Zn.  Our experiments performed on multiple length scales show that both compositions of Zn-barlowite ($x=0.95$ and $x \approx 0.5$) have ideal, fully occupied kagome layers of \ce{Cu^2+} cations, as does herbertsmithite.\cite{Freedman2010}  Fits to EXAFS data show no indications that Zn should occupy the kagome site but cannot definitively exclude this possibility. Single crystal anomalous diffraction allows us to determine that in the long-range structure no measurable amount of \ce{Zn^2+} occupies the kagome site. First-principles simulations and experimental XANES data, which probe the local structure, further confirm that it is energetically unfavorable for \ce{Zn^2+} to occupy a kagome site in both Zn-barlowite and herbertsmithite, and simulated models with \ce{Zn^2+} on the kagome site are not consistent with the data. This provides strong evidence supporting our claim that Zn-substituted barlowite with $x \gtrsim 0.5$ is a viable QSL candidate,\cite{Smaha2020} which has also been supported by a recent $\mu$SR study.\cite{Tustain2020}

Next, we are able to make site assignments for the two distinct interlayer sites observed in both compositions of Zn-substituted barlowite---$x \approx 0.5$ (\textbf{Zn$^\mathrm{D}_{0.52}$} and \textbf{Zn$^\mathrm{H}_{0.56}$}) and $x=0.95$, as elucidated by ICP-AES---by SCXRD measurements in this and previous work.\cite{Smaha2020}  As shown in Figure \ref{fgr:schematic}, one site lies at the center of the trigonal prismatic coordination environment, and the second lies off-center---matching the sites seen in the barlowite parent compound. The relative occupancies of these two sites, as measured by SCXRD and PXRD, differ as the composition changes. As the amount of Zn substitution increases, the occupancy of the centered site increases, suggesting that it is occupied by \ce{Zn^2+}. This is consistent with predictions from Jahn-Teller theory, but these standard crystallographic measurements cannot, however, provide direct evidence for which sites Zn and Cu occupy in the structure. We use X-ray anomalous diffraction to show that \ce{Zn^2+} occupies the centered interlayer site while \ce{Cu^2+} occupies the kagome and off-center interlayer sites. Two sites are observed even for the single crystal found in the bulk \textbf{Zn$^\mathrm{D}_{0.95}$} sample, which is likely not representative of the bulk and whose occupancies for the centered and off-center sites are 0.85 and 0.15, respectively. This conclusion is supported by the nearly identical Zn L-edge XANES spectra (which reflect the local structure) of \textbf{Zn$^\mathrm{H}_{0.56}$} and \textbf{Zn$^\mathrm{H}_{0.95}$}, implying that \ce{Zn^2+} must occupy the same site in both compositions of Zn-substituted barlowite. 

However, herbertsmithite contains one interlayer site occupied by a mixture of both \ce{Zn^2+} and \ce{Cu^2+}.\cite{Freedman2010}  The composition of the typical herbertsmithite crystal mirrors that of the \textbf{Zn$^\mathrm{D}_{0.95}$} crystal, but its site splitting behavior has never been observed in herbertsmithite. This points to a fundamental difference in the interlayer motifs between the two compounds.

In particular, we explore differences in the interlayer \ce{Zn^2+} coordination between Zn-substituted barlowite and herbertsmithite, showing that they are nearly indistinguishable on one length scale but strikingly distinct when considered at other length scales. Their similarity on a local scale is exemplified by the near equivalence of their Zn K-edge EXAFS spectra (Figure \ref{fgr:exafs}C), which occurs because \ce{Zn^2+} in both compounds occupies a centered site with six equivalent Zn--O bonds. However, on a slightly longer---but still local---length scale the differences between these structures become apparent. A combination of high-resolution XANES measurements and first-principles simulations reveal a handle for distinguishing (pseudo-)octahedral and trigonal prismatic \ce{Zn^2+} coordination environments by the number of peaks close to the edge jump, and this trend is borne out by experimental data and calculations of other materials that share these \ce{Zn^2+} coordination environments.  

Taken together, our observations suggest that Zn-barlowite is less disordered than herbertsmithite due to its unique interlayer site splitting (mixed occupancy is not observed on either interlayer site). This may also allow it to support a higher amount of Zn substitution, thus facilitating more optimal low-energy measurements of its likely QSL ground state. Such measurements in herbertsmithite single crystal samples are made difficult due to the presence of magnetic \ce{Cu^2+} ``impurities" between the kagome layers.\cite{Han2016b} Such detailed neutron scattering measurements require deuterated single crystals, which we report here for the first time. Additionally, the significant difference between the two QSL candidates in the free energies of the ``kagome Zn" and ``interlayer Zn" models suggests that Zn-substituted barlowite is more thermodynamically stable than herbertsmithite, so large crystals may be grown with the highest levels of Zn-substitution. This is consistent with previous first-principles calculations.\cite{Liu2015a} Interestingly, the similarity of the bulk magnetic susceptibility of Zn-barlowite and herbertsmithite, in spite of the measured differences in the local structure around the interlayer defects, indicates that the kagome QSL is robust to this type of disorder.\cite{Smaha2020} Hence, Zn-substituted barlowite, as a QSL candidate, has the potential to further advance the experimental realization of this exotic phase of matter.

\section{\label{sec:DataAvilability}Data and Materials Availability}
The data that supports the findings of this study is available from the corresponding authors upon reasonable request. Crystallographic Information Files (CIFs) have been deposited in the Cambridge Crystallographic Data Center (CCDC): 1899246, 1899248, 1995564.

%

\begin{acknowledgments}
The authors would like to thank S. Conradson and L.B. Gee for helpful discussions relating to EXAFS and S. Lapidus for assistance at APS beamline 11-BM. The work at Stanford and SLAC was supported by the U.S. Department of Energy (DOE), Office of Science, Basic Energy Sciences (BES), Materials Sciences and Engineering Division, under Contract No. DE-AC02-76SF00515. The experimental work and analysis was performed at the Stanford Institute for Materials and Energy Sciences (SIMES), and the computational work was performed at the Theory Institute for Materials and Energy Spectroscopies (TIMES). This research used resources of the Advanced Light Source, which is a DOE Office of Science User Facility under contract No. DE-AC02-05CH11231. Use of the Stanford Synchrotron Radiation Lightsource, SLAC National Accelerator Laboratory, is supported by the DOE, Office of Science, BES, under Contract No. DE-AC02-76SF00515. Use of the Advanced Photon Source, an Office of Science User Facility operated for the U.S. DOE Office of Science by Argonne National Laboratory, was supported by the U.S. DOE under Contract No. DE-AC02-06CH11357. NSF's ChemMatCARS Sector 15 is supported by the Divisions of Chemistry (CHE) and Materials Research (DMR), National Science Foundation, under grant number NSF/CHE-1834750. This research used resources of the National Energy Research Scientific Computing Center (NERSC), a DOE Office of Science User Facility operated under Contract No.  DE-AC02-05CH11231. Part of this work was performed at the Stanford Nano Shared Facilities (SNSF), supported by the NSF under award ECCS-2026822.  R.W.S. was supported by the Department of Defense (DoD) through the NDSEG Fellowship Program and by a NSF Graduate Research Fellowship (DGE-1656518). Certain commercial equipment, instruments, or materials are identified in this paper in order to specify the experimental procedure adequately.  Such identification is not intended to imply recommendation or endorsement by NIST, nor is it intended to imply that the materials or equipment identified are necessarily the best available for the purpose.
\end{acknowledgments}

\section{\label{sec:Contributions}Author Contributions}
R.W.S. and Y.S.L. conceived the study, interpreted the data, and wrote the manuscript with contributions and comments from all authors. R.W.S. synthesized barlowite \textbf{2} and Zn-barlowite and performed and analyzed X-ray diffraction measurements.  I.B. performed and analyzed simulations with C.D.P., J.V, and T.P.D. C.J.T. performed and analyzed EXAFS and XANES measurements. J.M.J. performed and analyzed anomalous diffraction measurements. J.P.S. and W.H. synthesized herbertsmithite and barlowite \textbf{1}, respectively. S.G.W., Y.-S.C., and S.J.T aided at beamtimes.
\end{document}